\documentclass[twocolumn]{aastex63}

\usepackage{multirow}
\usepackage{hyperref}

\begin{document}
	
	\title{Detectability of Artificial Lights from Proxima b}
	\author{Elisa Tabor}
	\affiliation{Department of Physics, Stanford University,  452 Lomita Mall, Stanford, CA 94305}
	\email{etabor@stanford.edu}
	\author{Abraham Loeb}
	\affiliation{Department of Astronomy, Harvard University, 60 Garden St., Cambridge, MA 02138}
	\email{aloeb@cfa.harvard.edu}
	
	\begin{abstract}
		We investigate the possibility of detecting artificial lights from Proxima b's dark side by computing light curves from the planet and its host star. The two different scenarios we consider are artificial illumination with the same spectrum as commonly used LEDs on Earth, and a narrower spectrum which leads to the same proportion of light as the total artificial illumination on Earth. We find that the James Webb Space Telescope (JWST) will be able to detect LED type artificial lights making up 5\% of stellar power with 85\% confidence, assuming photon-limited precision. In order for JWST to detect the current level of artificial illumination on Earth, the spectral band must be $10^3$ times narrower. Our predictions require optimal performance from the NIRSpec instrument, and even if not possible with JWST, future observatories like LUVOIR might be able to detect this artificial illumination.
		
		\vspace{4mm}
	\end{abstract}
	
	%	\keywords{}
	
	\section{Introduction}
	
	Proxima b is one of the best targets outside our solar system in the search for extraterrestrial life \citep{Anglada2016, Ribas2016, Turbet2016, Berdyugina2019}. It resides in the habitable zone of its star, suggesting liquid water could exist at its surface; it is a similar mass to Earth, and it orbits the nearest star to the sun, 4.2 light years away, so we can reasonably expect to characterize it in the near future. The important question is then whether Proxima b can sustain intelligent life and how to go about detecting it \citep{Turbet2016, Kreidberg2016, Lingam2017}.
	
	Since the discovery of Proxima b five years ago, there have been many searches for life on the planet \citep{Anglada2016, Ribas2016, Turbet2016}, culminating most recently with a tentative detection of a radio signal from BLC-1 originating from its direction (Breakthrough Listen, submitted 2021). Previous studies involved estimating the water content on the planet, projecting the rate at which Proxima b might lose its atmosphere due to stellar flares, reconstructing the evolution of the radius and luminosity of the planet to find its current rotation rates, and studies of its climate and atmosphere \citep{Turbet2016, Ribas2016, Kreidberg2016}. A generic aspect of a technological civilization is the production of artificial light. In this \textit{Letter}, we examine the detectability of artificial lights originating from Proxima b.
	
	Owing to its proximity to the star, Proxima b is likely to be tidally-locked with a permanent dayside and nightside. This exacerbates the need for artificial illumination of the nightside for it to be attractive for technological habitability, as it is otherwise permanently dark.
	
	The lightcurves of Proxima b involve several factors, including the radius of the planet ($\sim 1.3$ R$_{\oplus}$ for Proxima b) and host star ($\sim 0.14$ R$_{\odot}$ for Proxima Centauri), orbital period (11 days), orbital semi-major axis ($\sim$ 0.05 AU), albedo ($\sim$ 0.1 if analogous to the Moon), and orbital inclination \citep{Anglada2016}.
	
	To estimate the inclination of Proxima b's orbit, we use information about Proxima c \citep{Damasso2020}. This planet orbits at 1.5 AU, so it remains outside of Proxima Centauri’s habitable zone. Proxima c's large semi-major axis has allowed to find its orbital inclination, and the HARPS and UVES spectrograph obtained an orbital inclination of $i = 152\pm 14$ degrees \citep{Kervella2020}. Since the variation of orbital inclinations in the solar system is about $\pm7\%$, we use an inclination of $i = 2.65\pm 0.43$ radians for Proxima b.
	
	We consider the James Webb Space Telescope (JWST) to assess the feasibility of detecting artificial lights on Proxima b \citep{Beichman2014}. JWST will allow to characterize the atmosphere of Proxima B and find out how much energy transport occurs on the planet \citep{Kreidberg2016, Turbet2016}. Here we perform a detailed calculation of the lightcurves from Proxima b and simulate signal-to-noise calculations using the JWST Exposure Time Calculator (\href{https://jwst.etc.stsci.edu/}{ETC}).
	
	The organization of this \textit{Letter} is as follows. In section 2 we describe our methods for calculating lightcurves and our error analysis. Our results are described in section 3. Finally, we discuss the implications of these findings in section 4.

	\section{Methods}
	
	\subsection{Proxima b Lightcurves}
	
	We calculate the lightcurves from Proxima b using the Exoplanet Analytic Reflected Lightcurves (EARL)  \citep{Haggard2018}. The uniform albedo map corresponds to the $Y_0^0$ spherical harmonic and the flux equation,
	\begin{equation}\label{F00}
		F_0^0=\frac{1}{3\pi^{3/2}}(\sin w-w\cos w),
	\end{equation}
	where $w$ is the width of a lune cut out by the great circles defined by the subobserver and substellar points. The value of $w$ ranges from 0 to $\pi$. We must also multiply the reflected light by the albedo $A$, for which we make the more conservative approximation of $A=0.1$, the albedo for the Moon and for typical rocky bodies \citep{Usui2013}.
	
	\begin{figure}[h]
		\epsscale{1.15}
		\plotone{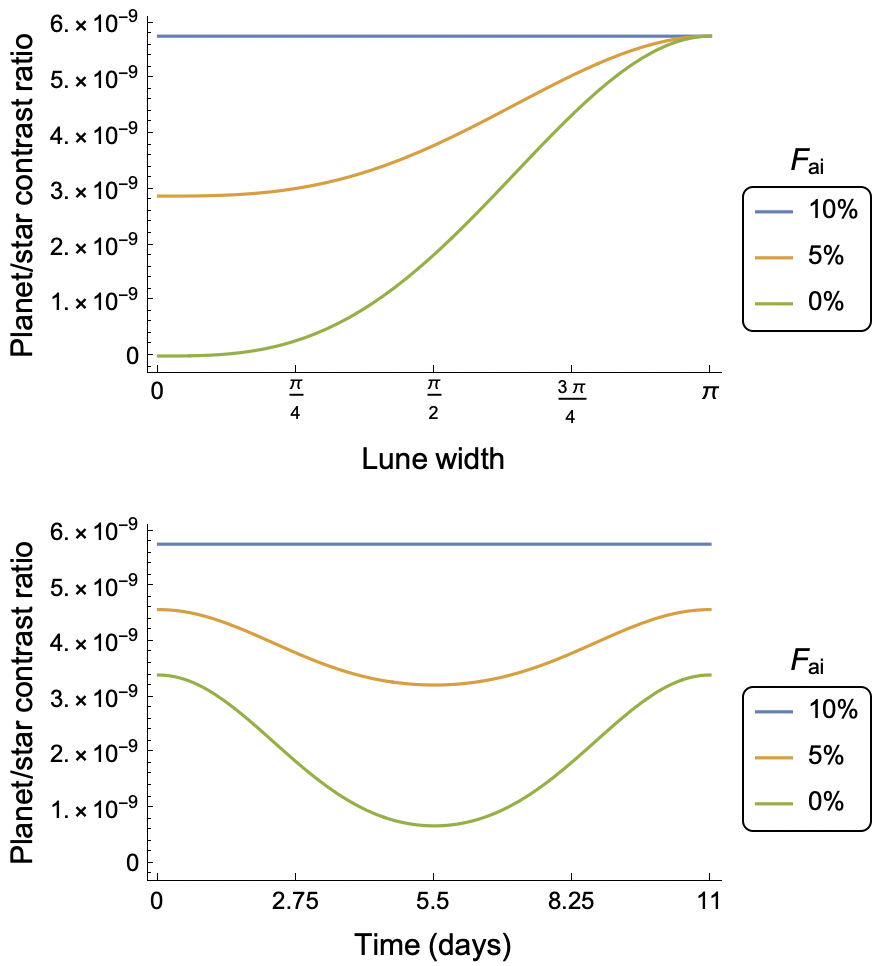}
		\caption{The lightcurves from Proxima b calculated using EARL, with three different coefficients $F_{ai}$ representing the percentage of stellar power being illuminated on the dark side of the planet. The blue curve represents $F_{ai}=0.1$, which equals the value we assume for the albedo. Thus the amount of artificial illumination on the night side is equal to the amount of light reflected from the day side. The green curve, for $F_{ai}=0$, represents no artificial illumination, so the night side is fully dark. \textit{Top panel}: the planet to star ratio depends solely on the lune width. \textit{Bottom panel}: the ratio depends on time (in days), orbital angular frequency, and inclination.}
		\centering
		\label{fig:lightcurves}
		%		\vspace{3mm}
	\end{figure}
	
	The lightcurve calculated with EARL only shows the light reflected from Proxima b, so to simulate artificial light in addition to the reflected light, we introduce a free variable to equation (\ref{F00}), $F_{ai}$. We assume the artificial light illuminates the night side of the planet facing away from its host star, so $F_{ai}$ represents the proportion of artificial light coming from Proxima b's dark side relative to the reflected starlight on the daylight hemisphere. Since the maximum value $w$ can take is $\pi$, and $F_0^0(\pi) = \frac{\pi}{3\pi^{3/2}}$, in order to purely consider the dark side, we examine $\frac{\pi}{3\pi^{3/2}}-F_0^0$. Thus, our new flux equation is
	$$F=\frac{A}{3\pi^{3/2}}(\sin w-w\cos w) + \frac{F_{ai}}{3\pi^{3/2}}(\pi-\sin w+w\cos w).$$
	%	where $F_{ai}$ represents the percentage of stellar power being illuminated on the dark side of the planet.
	To obtain the desired flux, we multiply our new $F$ by 
	$$\frac{\pi R^2}{4\pi d^2}$$
	where $R$ is the radius of Proxima b and $d$ is the distance between Proxima b and its star. The top panel of Figure \ref{fig:lightcurves} shows the flux dependence on lune width for different values of $F_{ai}$. The blue curve represents $F_{ai}=1$, which occurs when the night side is fully illuminated to the same brightness as the day side facing Proxima Centauri. The green curve, showing $F_{ai}=0$, represents no artificial illumination, in which case the night side remains fully dark.
	
	We next change coordinates from lune width to a dependence on time, inclination, and orbital angular frequency using the equations in Appendix C of \cite{Haggard2018}. Using the inclination of Proxima c of 2.65 radians and an orbital angular frequency of $(2\pi/11)$days, we obtain the curves in the bottom panel of Figure \ref{fig:lightcurves}.  The minimum for the green curve in the bottom panel is greater than 0 since Proxima b doesn't transit, so a nonzero fraction of the dayside is always visible \citep{Jenkins2019}. The flux equations give unitless ratios, so we must multiply the reflected light by the stellar spectrum and the artificial light by the predicted spectrum of the artificial illumination on Proxima b.
	
	We consider two possibilities for the artificial spectrum: one with commonly used LEDs, and another with the same $F_{ai}$ as currently used on Earth ($\sim 10^{-4}$). These spectra correspond to a Gaussian distribution centered at 1.2 $\mu$m (the peak of Proxima Centauri's light spectrum) with variance 0.12 $\mu$m and $10^{-4} \mu$m, respectively. We normalized the Gaussians so that when $F_{ai}=A$, the integral of the LED spectra equals the integral of reflected light from the host star. The variation in the width of these Gaussians allows us to examine the impact of more focused lights on their detectability.
	
	Finally, to calculate the spectral radiance density per unit frequency of Proxiam Centauri and Proxima b, we use the blackbody form,
	$$B(\nu) = \frac{2 h\nu^3}{c^2}\frac{1}{e^{h\nu/(k_B T)} - 1}$$
	with $T=250$K for the planet and $T=3000$K for the host star \citep{Kreidberg2016}. The final lightcurves are the sum of the spectral radiance densities of the star and of the planet, the reflected light from the star, and the artificial illumination of Proxima b.
	
	\subsection{JWST/NIRSpec}
	
	The JWST Exposure Time Calculator (ETC, available at \href{jwst.etc.stsci.edu}{jwst.etc.stsci.edu}) allows us to estimate the feasibility of detecting different values of $F_{ai}$. Since the peak of the stellar flux is around 1.5 microns, we work with NIRSpec's G140M/F100LP disperser-filter combination, which has wavelength range 0.97–1.84 microns and a spectral resolution of $R \sim 1000$.\footnote{\href{https://jwst-docs.stsci.edu/near-infrared-spectrograph/nirspec-observing-modes/nirspec-multi-object-spectroscopy}{jwst-docs.stsci.edu}} We set up using the spectral energy distribution of a blackbody at 3000K, and renormalize the source flux density to 4.384 in the K-band. The ETC produces a graph of the estimated number of photoelectrons per second for different wavelengths, which we use to calculate the predicted errors. The noise corresponds to the square root of the number of electrons $N_e$, so the noise-to-signal ratio is 1/$\sqrt{N_e}$ where $N_e$ is the number of electrons per second $n_e$ times the integration time $t_i$. Integrating the photons over one period (11 days) gives us an estimated error of $\frac{1}{\sqrt{N_e}} = \frac{1}{\sqrt{n_e*t_i}}.$
	
	\begin{figure}[h]
		\epsscale{1.15}
		\plotone{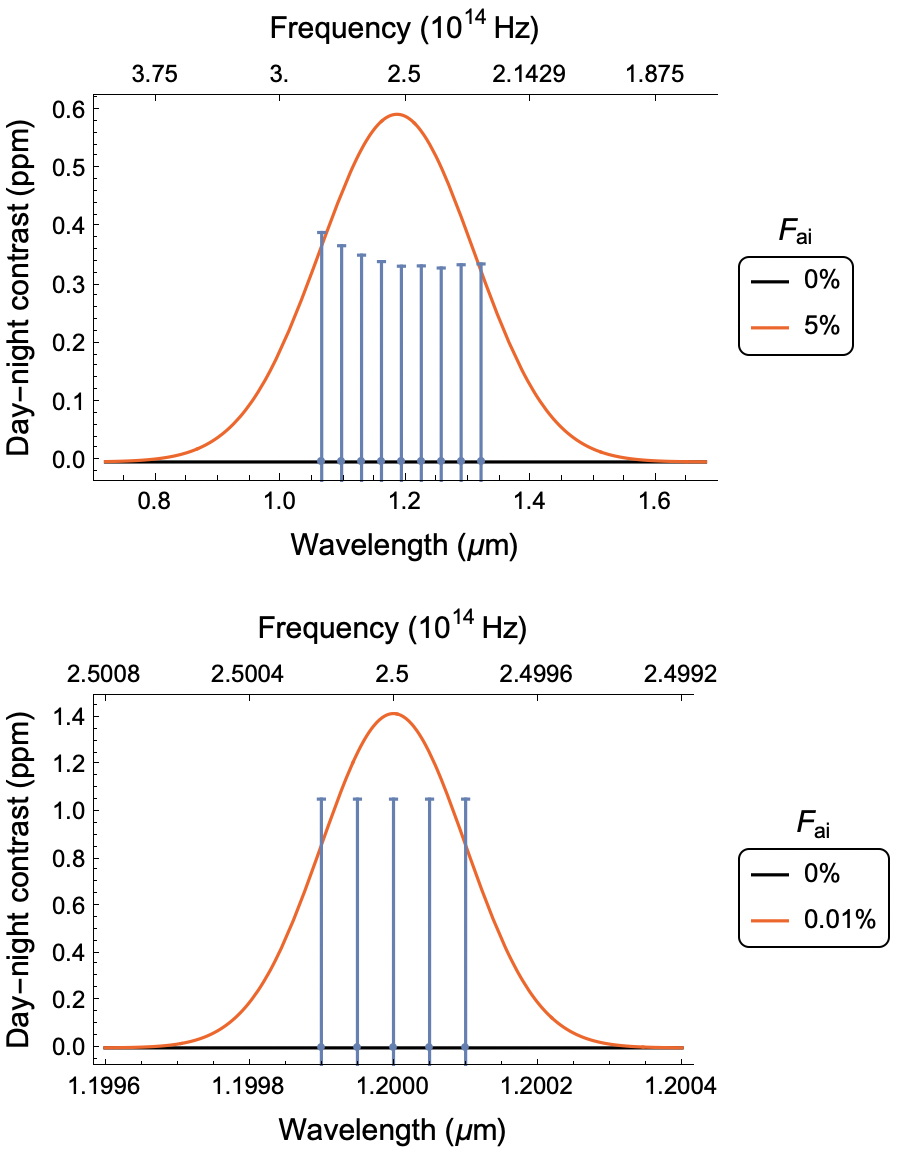}
		\caption{Difference between the measured spectrum at phase 0 and phase 0.5 in ppm. The black line involves the sum of the spectral radiance densities of Proxima b and its star and the reflected light from the star, while the red line adds the artificial illumination of Proxima b. The top axis shows the frequency (in $10^{14}$Hz) while the bottom axis shows the wavelength (in $\mu$m). The uncertainties are based on the photon noise from NIRSpec measurement predictions. \textit{Top panel}: the results obtained when the artificial light corresponds to commonly used LEDs on Earth. \textit{Bottom panel}: the curve which leads to the same $F_{ai}$ as currently used on Earth.}
		\centering
		\label{fig:curves}
%		\vspace{3mm}
	\end{figure}

	\section{Results}
	
	We now combine our error analysis with our calculated lightcurves to obtain full predictions of the lightcurves detected from Proxima b. Here we show the contrast between the light coming from the system at the start and in the middle of an orbital period.
	
	The top panel of Figure \ref{fig:curves} shows the results we obtained when the artificial light corresponds to commonly used LEDs on Earth, and the bottom panel shows the curve which leads to the same $F_{ai}$ as currently used on Earth, $\sim 10^{-4}$.
	
	\section{Discussion and Conclusions}
	
\begin{figure}
		\epsscale{1.18}
		\plotone{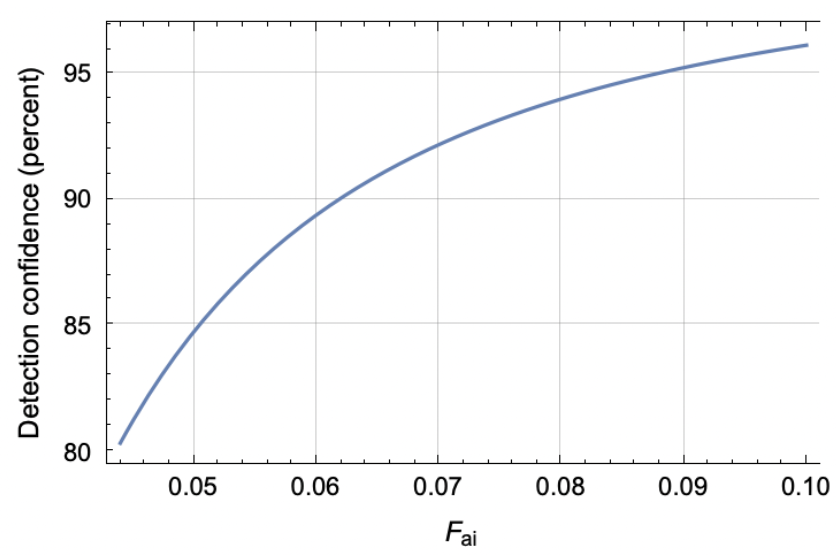}
		\caption{Chi-squared analysis of data for standard LEDs with width 0.12 $\mu$m.}
		\centering
		\label{fig:chi}
		\vspace{5mm}
	\end{figure}

	We use a reduced chi-square statistic to analyze the confidence of detecting certain values of $F_{ai}$, shown in Figure \ref{fig:chi}. For the LED spectrum, we find that JWST will be able to rule out values of $F_{ai}>5\%$ with 85\% confidence, and $F_{ai}>9\%$ with 95\% confidence. In order for JWST to detect a coefficient of $F_{ai}>0.001\%$ (the current value on Earth),with 85\% confidence, the spectrum of artificial illumination must be $10^3$ times narrower in frequency. In either case, JWST will thus allow us to narrow down the type of artificial illumination being used.
	
	Proxima b is tidally locked if its orbit has an eccentricity below 0.06, where for reference, the eccentricity of the Earth's orbit is 0.017 \citep{Ribas2016}. If Proxima b has a permanent day and nightside, the civilization might illuminate the nightside using mirrors launched into orbit or placed at strategic points \citep{Korpela2015}. In that case, the lights shining onto the permanent nightside should be extremely powerful, and thus more likely to be detected with JWST.
	
	%	If the eccentricity is greater than 0.06, the planet will be locked into a 3:2 spin-orbit resonance similar to Mercury \citep{Ribas2016}.
	
	In summary, we have simulated lightcurves from Proxima b and compared curves corresponding to the reflected stellar spectrum to curves with artificial lights corresponding to a narrower spectrum such as for LEDs. We have found that JWST will be able to show the existence of artificial illumination for standard LEDs 500 times more powerful than those currently found on Earth's, and for artificial illumination of similar magnitude to Earth's for a spectrum $10^3$ times narrower in frequency.
	
	Future extensions of this work could examine the sensitivity of the Large UV Optical Infrared Surveyor (LUVOIR), a telescope in the works with the goal of being launched in 2035 \citep{LUVOIR2019}. LUVOIR will allow us to confirm the artificial illumination, or lack thereof, with a higher degree of precision. Although the noise floor of JWST is not known, current estimates are at about 10 ppm \citep{Schlawin2021}. Detecting an effect of 1 ppm will require pushing the NIRSpec instrument well beyond its expected performance. Even if JWST is not able to detect artificial illumination on Proxima b, LUVOIR and other such telescopes may have significantly improved performance.
	
	\acknowledgments This work was supported in part by the Stanford Physics Department (for E.T.) and in part by a grant from the Breakthrough Prize Foundation (for A.L.). We thank Laura Kreidberg for insightful comments on an early draft of the paper.
	
%	\newpage
	\bibliographystyle{aasjournal}
	\bibliography{ProximaB}
	
\end{document}